\documentclass[8.5pt,twoside,twocolumn]{article}

\usepackage[super,sort&compress,comma]{natbib} 
\usepackage{mhchem}
\usepackage{times,mathptmx}
\usepackage{sectsty}
\usepackage{balance} 
\usepackage{natbib}
\usepackage{hyperref}
\hypersetup{colorlinks=true,linkcolor=blue,citecolor=blue,urlcolor=blue}
\usepackage{graphicx} 
\usepackage{lastpage}
\usepackage[format=plain,justification=raggedright,singlelinecheck=false,font=small,labelfont=bf,labelsep=space]{caption} 
\usepackage{fancyhdr}
\begin{document}

\thispagestyle{plain}
\fancypagestyle{plain}{
\renewcommand{\headrulewidth}{1pt}}
\renewcommand{\thefootnote}{\fnsymbol{footnote}}
\renewcommand\footnoterule{\vspace*{1pt}%
\hrule width 3.4in height 0.4pt \vspace*{5pt}} 
\setcounter{secnumdepth}{5}

\makeatletter 
\def\subsubsection{\@startsection{subsubsection}{3}{10pt}{-1.25ex plus -1ex minus -.1ex}{0ex plus 0ex}{\normalsize\bf}} 
\def\paragraph{\@startsection{paragraph}{4}{10pt}{-1.25ex plus -1ex minus -.1ex}{0ex plus 0ex}{\normalsize\textit}} 
\renewcommand\@biblabel[1]{#1}            
\renewcommand\@makefntext[1]%
{\noindent\makebox[0pt][r]{\@thefnmark\,}#1}
\makeatother 
\renewcommand{\figurename}{\small{Fig.}~}
\sectionfont{\large}
\subsectionfont{\normalsize} 

\fancyfoot{}
\fancyfoot[RO]{\footnotesize{\sffamily{1--\pageref{LastPage} ~\textbar  \hspace{2pt}\thepage}}}
\fancyfoot[LE]{\footnotesize{\sffamily{\thepage~\textbar\hspace{3.45cm} 1--\pageref{LastPage}}}}
\fancyhead{}
\renewcommand{\headrulewidth}{1pt} 
\renewcommand{\footrulewidth}{1pt}
\setlength{\arrayrulewidth}{1pt}
\setlength{\columnsep}{6.5mm}
\setlength\bibsep{1pt}


\twocolumn[
\begin{@twocolumnfalse}
\noindent\LARGE{\textbf{Unconventional Strain-Dependent Conductance Oscillations in Pristine Phosphorene}}
\vspace{7pt}

\noindent\large{\textbf{S. J. Ray$^{\ast}$\textit{$^{a}$,}}}
\noindent\large{\textbf{M. Venkata Kamalakar\textit{$^{b}$}}}

\vspace{0.6cm}



\noindent \normalsize{Phosphorene is a single elemental two-dimensional semiconductor that has quickly emerged as a high mobility material for transistors and optoelectronic devices. In addition, being a 2D material, it can sustain high levels of strain, enabling sensitive modification of its electronic properties. In this paper, we investigate the strain dependent electrical properties of phosphorene nanocrystals. Performing extensive calculations we determine electrical conductance as a function uniaxial as well as biaxial strain stimulus, and uncover a unique zone phase diagram. This enables us to uncover for the first time conductance oscillations in pristine phosphorene, by simple application of strain. We show that how such unconventional current-voltage behaviour is tuneable by the nature of strain, and how an additional gate voltage can modulate amplitude (peak to valley ratio) of the observed phenomena and its switching efficiency. Furthermore, we show that the switching is highly robust against doping and defects. Our detailed results present new leads for innovations in strain based gauging and high-frequency nanoelectronic switches of phosphorene.}
\vspace{0.5cm}
\end{@twocolumnfalse}
]

\footnotetext{\textit{$^{a,*}$Department of Physics, Indian Institute of Technology Patna, Bihta 801106, India; E-mail: ray@iitp.ac.in; ray.sjr@gmail.com}}

\footnotetext{\textit{$^{b}$Department of Physics and Astronomy, Uppsala University, PO Box 516, 75120, Uppsala, Sweden}}



\section{Introduction}
\begin{figure*}
\begin{center}
\includegraphics[width=14cm]{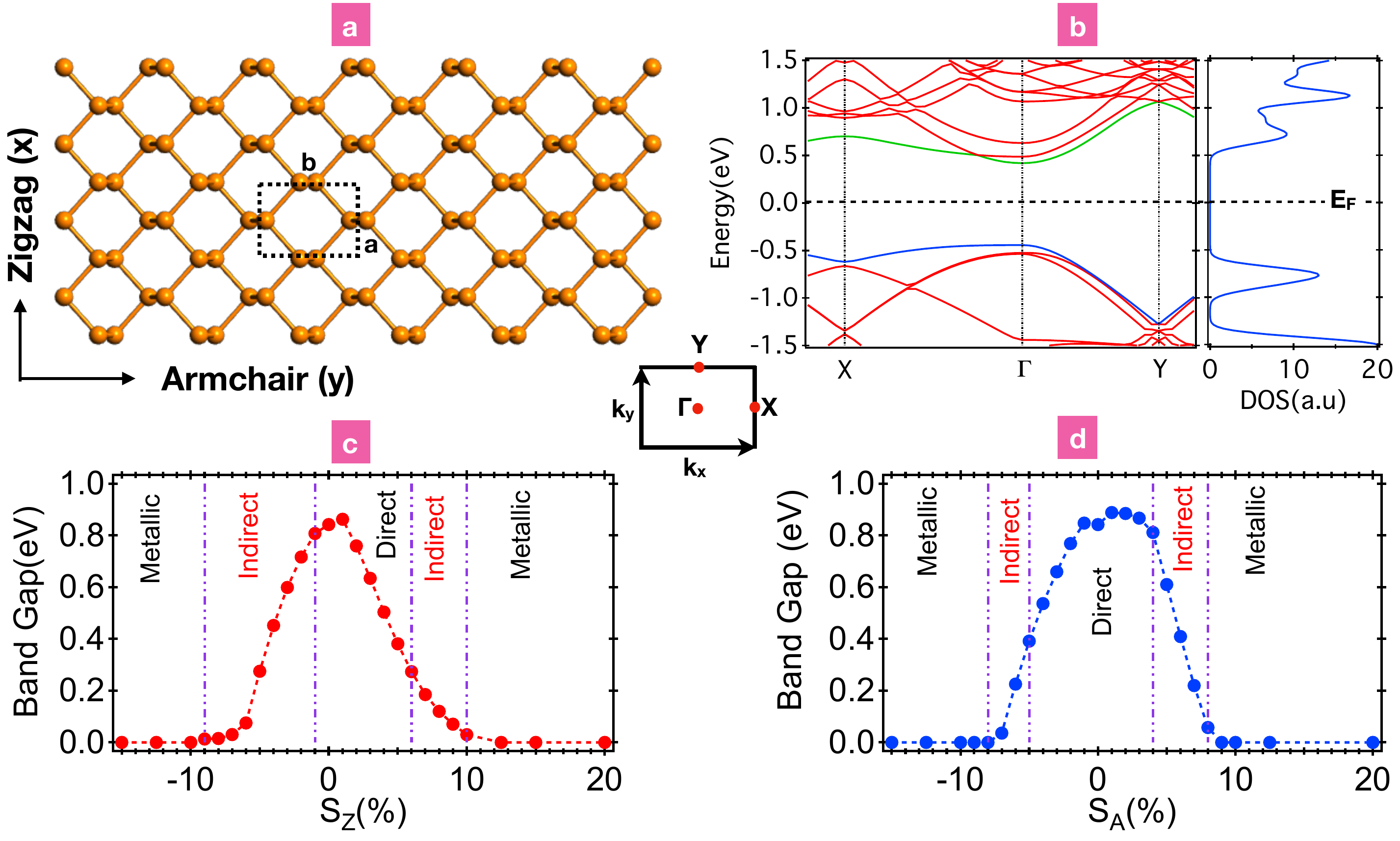}
\caption{{\small (a) Schematic of the monolayer Phosphorene (top view) with the unit cell marked through dotted rectangle, (b) Band Structure of a $3\times3$ supercell of Phosphorene at $S_Z$ = 1\% and corresponding density of states (right), (c) Dependence of electronic band gap as function of uniaxial strain(s) along the armchair and (d) zigzag direction.}}
\label{fig.1}
\end{center}
\end{figure*} 
Over the last one decade the two-dimensional (2D) crystals became novel testbeds for exploring fundamental phenomena in condensed matter physics, in addition to their demonstrated potential for electronic, spintronic and optoelectronic applications \cite{NovoselovPNAS}. Out of an increasing number of explored 2D materials, Phosphorene, the atomically thin crystal of black phosphorus quickly rose to prominence for possessing striking physical attributes for multiple applications. Firstly, why Phosphorene or black Phosphorene is so exceptional? So far, it has been the single reliable 2D semiconductor, that shows a high ON/OFF ratios \cite{Li2014, Liu2014}, ambipolar transport \cite{Das2014, Venkata2015}, a high mobility \cite{Long2016} comparable to that of silicon and GaAs. Black Phosphorus is also the only 2D semiconductor, where possibility of planer spin transport has been indicated \cite{Kamalakar2015} and realised \cite{Avsar2017} compared to length scales observed in graphene. Owing to these exhibits, research in Phosphorene is more intense than ever, with new phenomena and applications continuing to fascinate scientists and technologists.

From the point of mechanical properties, 2D crystals truly standout when compared to thin films of conventional materials made by top-down approaches. These materials possess outstanding flexibility to sustain strain levels upto 10 times more than bulk materials, which makes them special and opens up new possibilities to explore and operate with extremely high flexibility \cite{Rafael2015, Wei2014}. Strain introduces changes in the band structure such as modifications in bandgap and effective mass of carriers, enabling engineering of 2D materials. Several studies have been carried out to see the effect of strain on Phosphorene such as phase transition \cite{Rodin2014, Guan2014, Wu2015, Elahi2015, Peng2014, Han2014}, changing conduction anisotropy \cite{Fei2014, Rodin2014}, negative Poisson's ratio \cite{Jiang2014}, Dirac-like cones \cite{Can2015, fei2015}, bandgap engineering \cite{Guan2014, Han2014, Zhu2014, Deniz2014, Guo2014, Li2014c, Qiao2014, Rudenko2014, Tran2014, Dai2014, Cai2014} etc.

Here we observe a strain dependent unconventional and reversible conductance oscillation (CO) in phosphorene nanocrystals. We obtain a comprehensive strain dependent phase diagram of 2D Phosphorene illustrating different electronic phases as a function of both uniaxial/biaxial strains. For the first time, we uncover the existence of a strain induced conductance oscillations in the IV characteristic of Phosphorene tuneable with applied strain value. The primary region with the highest peak to valley ratio (PVR) can be tuned by an applied gate voltage to get more than 100\% enhancement in the switching efficiency (S$_E$). Furthermore, we show that such oscillation are highly robust, irrespective of doping or defect and can open up an avenue of new possibilities for Phosphorene in highly controllable 2D nanoelectronic strain based applications.

\section{Computational Procedure}

First-principles based Density Functional Theory (DFT) method was used to calculate the structural and electronic properties of Phosphorene under different strained conditions. The self-consistent calculations were performed using the Perdew-Burke-Ernzerhof (PBE) \cite{PBE} exchange-correlation functional under the Generalised Gradient Approximation (GGA) as implemented within the Atomistix Toolkit \cite{ATK}. The transport calculations were performed using the Non-equilibrium Green's function (NEGF) formalism combined with DFT \cite{Soler2002, Brandbyge2002} implemented within the same \cite{ATK}. For band structure calculation, the reciprocal space of the Brillouin Zone was sampled using a $16\times16\times1$ Monkhorst-Pack\cite{Monkhorst} $k$-mesh. Periodic boundary conditions were used on the unit cells of Phosphorene with Double-$\zeta$ polarised (DZP) basis sets with an energy cut-off limit of 180 Ry. To minimise interactions between neighbouring layers, a minimum vacuum space of 15$\rm{\AA}$ was used in the non-periodic directions. All the geometries were structurally optimised until the force on each atom was less than 10$^{-3}$ eV/\AA.

In the 2-probe geometry, the current was calculated using the Landau-B$\ddot{u}$ttiker formula\cite{Landauer, Butikker},
\begin{equation}
I = \frac{2e}{\hbar}\int_{-eV/2}^{eV/2}T(E,V)[f_L(E-\mu_L) - f_R(E-\mu_R)]dE
\label{eqn.1}
\end{equation}
where $f(E)$ is the Fermi distribution function and T(E,V) is the transmission function at a given energy E and applied bias voltage V and $\mu_L, \mu_R$ are the chemical potentials of the left and right electrodes. For the transport calculation, a $k$-point grid of  $1\times1\times300$ (300 in the transport direction) was used. A convergence study was performed  with respect to $k$-point sampling, where no changes in the transmission were observed upto 500 $k$-points along the transverse direction.  
\begin{figure*}
\begin{center}
\includegraphics[width=18cm]{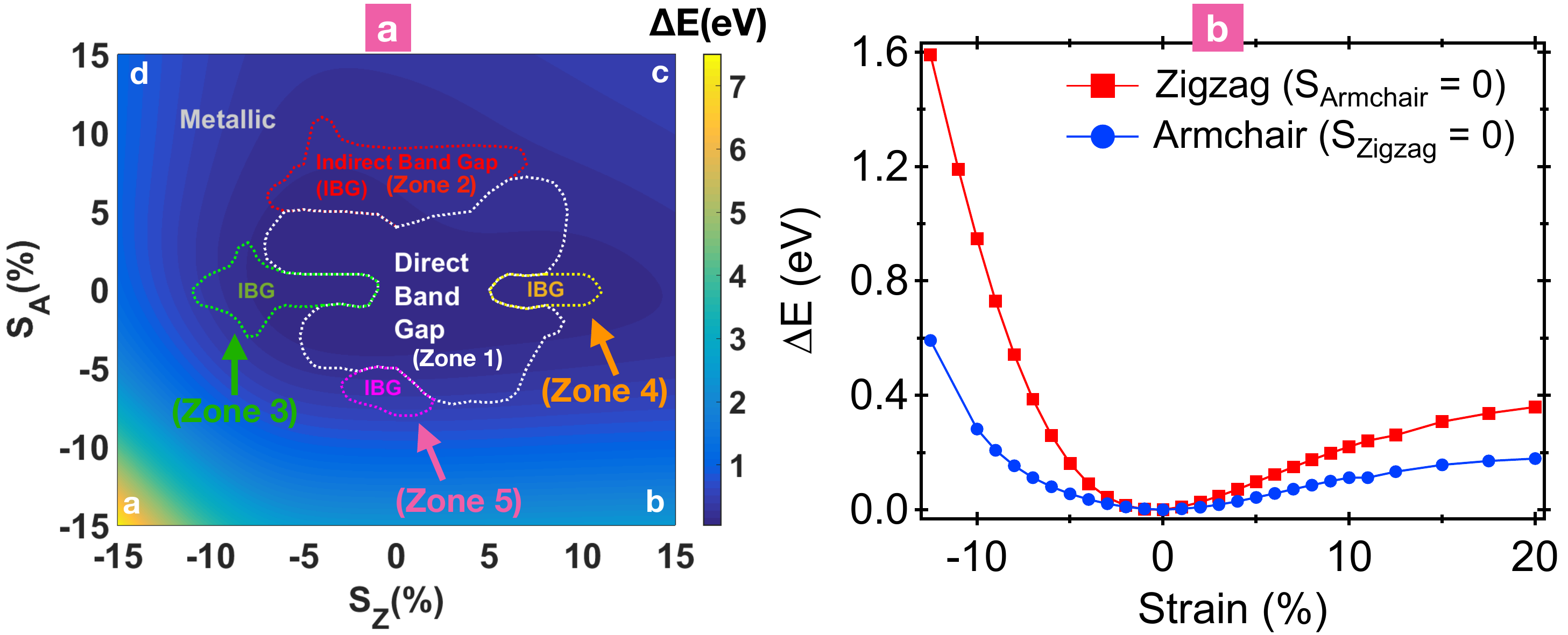}
\caption{{\small (a) Strain energy phase diagram of Phosphorene as a function of various strains along the armchair and zigzag directions. Different regions are labelled corresponding to different electronic phases of the material, (b) Dependence of the strain energy as a function of uniaxial strains along the armchair and zigzag directions.}}
\label{fig.2}
\end{center}
\end{figure*} 

\section{System Description}

Black Phosphorus (BP) is a stable allotrope of Phosphorus and single layer BP has a puckered honeycomb structure resulting from $sp^3$ hybridisation with each P atom covalently bonded with three neighbouring P atoms as depicted in Fig.\ref{fig.1}(a). The primitive unit cell of Phosphorene was obtained from our previous studies \cite {Ray2016a, Ray2016b}. For the band structure calculation, a $3\times3$ supercell of Phosphorene was used and this was later converted into a 2-probe configuration with semi-infinite electrodes for current-voltage calculation. 

The strain was applied by way of tuning the lattice parameters in the two planar directions : along x (zigzag) or y (armchair). Two types strains were considered in this study : (i) uniaxial strain (US) and biaxial strain (BS). For each of them, compressive strain (CS) and tensile strain  (TS) regions were studied. The applied uniaxial strains are defined as, $S_{Z} = (a - a_0)/a_0$, $S_{A} = (b - b_0)/b_0$ where $a(b)$ is the lattice constant at that given strain value and $a_0 (b_0)$ is the value under relaxed/unstrained condition along the zigzag (armchair) direction respectively. The positive value of strain indicates expansion (TS) and negative value signifies CS. It is to be noted that at a specific value of the US, the lattice constant in the transverse direction was allowed to fully relax to ensure the energy minimisation condition.

\section{Results and Discussion}

\subsection{Electronic Structure}
The electronic band structure (BS) of Phosphorene at 1\% US along the zigzag direction ($S_Z$ = 1\%) is illustrated in Fig.~\ref{fig.1}(b). The semiconducting nature can be confirmed from the BS and the density of states profile with a direct band gap ($E_g$) of $\sim$ 0.87 eV at the $\Gamma$ point. The curvatures of the conduction band (CB) and the valence band (VB) along the zigzag direction is much flatter compared to that along the armchair direction. This suggests a difference in the carrier effective mass ($1/m^* \sim \partial^2E/\partial k^2$), which both for the electron and holes along the $\Gamma\rightarrow X$ direction are significantly heavier than that along the $\Gamma\rightarrow Y$ direction. The hole effective mass ($m_h^*$) corresponding to  Fig.~\ref{fig.1}(b), is 5.68$m_0$ along the $\Gamma\rightarrow X$ and 0.15$m_0$ ($m_0$ is the rest mass of an electron) along the $\Gamma\rightarrow Y$ direction. It demonstrates the anisotropic nature of Phosphorene leading to dispersive band alignments along its two principal directions.

Dependence of the $E_g$ for various values of $S_Z$ and $S_A$ is shown in Fig.~\ref{fig.1}(c-d). On the application of TS along the zigzag direction (Fig.~\ref{fig.1}(c)) [$S_A = 0$], the $E_g$ of Phosphorene gets reduced with an increase in $S_Z$ and it becomes negligible above $S_Z =$10\%, beyond which the material becomes metallic. Between $S_Z =$ 1-10\%, the decay of the value of $E_g$ can be analytically described using a power law dependance with an average fall rate of $\sim$ 0.09 eV/\% strain. Similar behaviour was also observed for the CS region, where the power law decay resulted in a $E_g$ decay of $\sim$ 0.1 eV/\% strain. Strain dependent decay of $E_g$ has been reported for graphene \cite{Bhattacharya2011, Peng2011}, MoS$_2$ \cite{Gerd2015, Lloyd2016} etc, which for Phosphorene can be explained using its non-planar structure and varying degree of overlap between the $p$-orbitals as a function of strain. For -1\%$ < S_Z<$ 6\%, Phosphorene is a direct band gap (DBG) semiconductor and Indirect band gap (IBG) structure is observed on both sides of this region, for 6\% $<S_Z<$ 10\% and -9\% $<S_Z<$ -1\%. Beyond these limits ($S_Z<$ -9\%, $S_Z>$ 10\%), $E_g$ disappears and it becomes metallic. Similar phase transition is observed for various values of $S_A$ keeping $S_Z = 0$ as illustrated in Fig.~\ref{fig.1}(d). The DBG is observed for -5\% $<S_A<$ 4\% and IBG for 4\% $<S_A<$ 8\%, -8\% $<S_A<$ -5\%. The metallic phase is found for $S_Z >$ 8\% and $S_Z <$ -8\%. This indicates that tunability of the electronic properties of Phosphorene can be easily achieved by using external strain, which can be very useful in nanoelectronics. For example, the DBG region will be of interest for optical applications while the tunability of $E_g$ is more desired for transistor fabrication. 
 \begin{figure}
\begin{center}
\includegraphics[width=8cm]{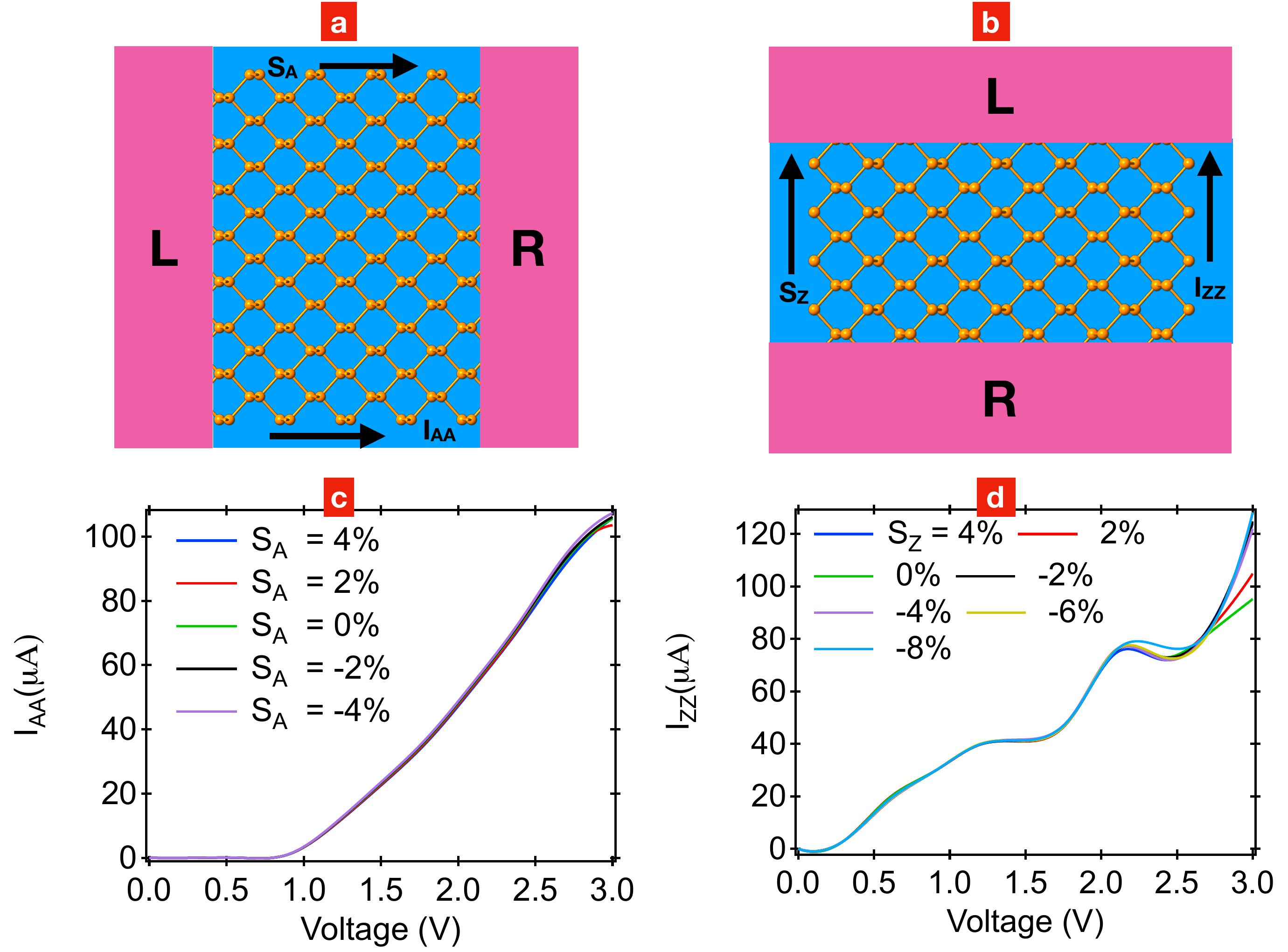}
\caption{{\small Two-probe conduction geometry in Phosphorene along the (a) armchair and (b) zigzag directions. Current-Voltage dependence for the above geometries for different values of uniaxial strains along the (c) armchair and (d) zigzag directions respectively.}}
\label{fig.3}
\end{center}
\end{figure} 
The lone pair orbital of the P-atoms lying out-of-plane of the BP layer are primarily of $p_z$ character and contribute to the bonding with foreign atoms or adsorbates. In our case, we considered monolayer with no adsorbates where the repulsion between interlayer lone-pairs is absent. Under the application of strain, the lone pair contributes by changing the character of the CBM from $p_z \rightarrow p_y$ type ($\sigma^* \rightarrow \pi^*$ state) which leads to a lowering of the band gap.

Further insight into the electronic properties of Phosphorene can be obtained from the 2D strain phase diagram of Phosphorene as illustrated in Fig.~\ref{fig.2}(a). The z-axis (colour bar) represents strain energy which is defined as,
\begin{equation}
\Delta E = (E_{s} - E_0)/N
\label{eqn.3}
\end{equation}
where $E_s (E_0)$ is the energy in strained (unstrained) configuration and $N$ is the number of unit cells. The dotted lines around various zones are estimated from the BS obtained at various values of $S_A$ and $S_Z$, which are plotted over the 2D strain energy profile to obtain the phase diagram. Around $S_A = S_Z = 0$, DBG is observed in Zone 1 for small values of strains around it. Four different zones of various sizes with IBG can be identified on four sides of this region. Moving from Zone 1, Zone 2 is found at a higher value of $S_A > 0$, while in the CS region Zone 5 is observed. Similarly Zone 3 and Zone 4 are lying along $S_A = 0$ line, for enhanced values of $S_Z$. For the rest of the phase diagram, the metallic phase of Phosphorene is observed. Such detailed illustration of the electronic phase diagram of Phosphorene is reported here for the first time and it can be immensely useful in choosing the right kind of region for the desired application by way of applying suitable strain. The critical strain for our case was estimated to be $\sim$ 30\% setting up the predicted elastic limit of operation in Phosphorene, which is also of similar order of magnitude as that of graphene and MoS$_2$ \cite{Lee2008, Castellanos2012}.

The line scan of $\Delta E$ (Eqn.~\ref{eqn.3}) for US as function of $S_A$ and $S_Z$ is shown in Fig.~\ref{fig.2}(b). For both the red and blue curves, the energy minimum is observed at the unstrained configurations. With an increase in the CS, the in-planar bond lengths (inter-atomic distances) get reduced, which provides a huge increase in the $\Delta E$ and higher degrees of overlap of $p_x, p_y$ orbitals of P-atoms. Compared to this, the increase in $\Delta E$ is relatively slower with an increase in the TS. In this case, $\Delta E$ increases sharply at first and shows a steady increase thereafter at higher values of TS.

The monotonic increase in $\Delta E$ around the unstrained configurations suggests that the system stays within its elastic limit within this range of applied strains and the deformed structure can return to its original unperturbed configuration on the removal of the strain. Over the entire range of strain, $\Delta E$ along the zigzag direction is higher than its armchair counterpart and the ratio is more than 2 at a high value of TS and CS. This suggests that the strain-deformation response in Phosphorene is different along the two directions and application of strain is difficult along the zigzag side compared to the armchair direction. This is due to the intrinsic structural anisotropy of Phosphorene arising out of unequal bond angles in its crystal structure. This can be related to the variations in $\Delta E$ at four corners ($a, b, c, d$) in Fig.~\ref{fig.2}(a). The highest value of $\Delta E$ is observed at $(a)$ and smallest at $(c)$ due to a simultaneous presence of CS (TS) along the two directions. $\Delta E$ at ($b$) and ($d$) are comparable in magnitude, but slightly higher at ($d$) as armchair is the preferred strain direction in Phosphorene.  
\begin{figure}
\begin{center}
\includegraphics[width=9cm]{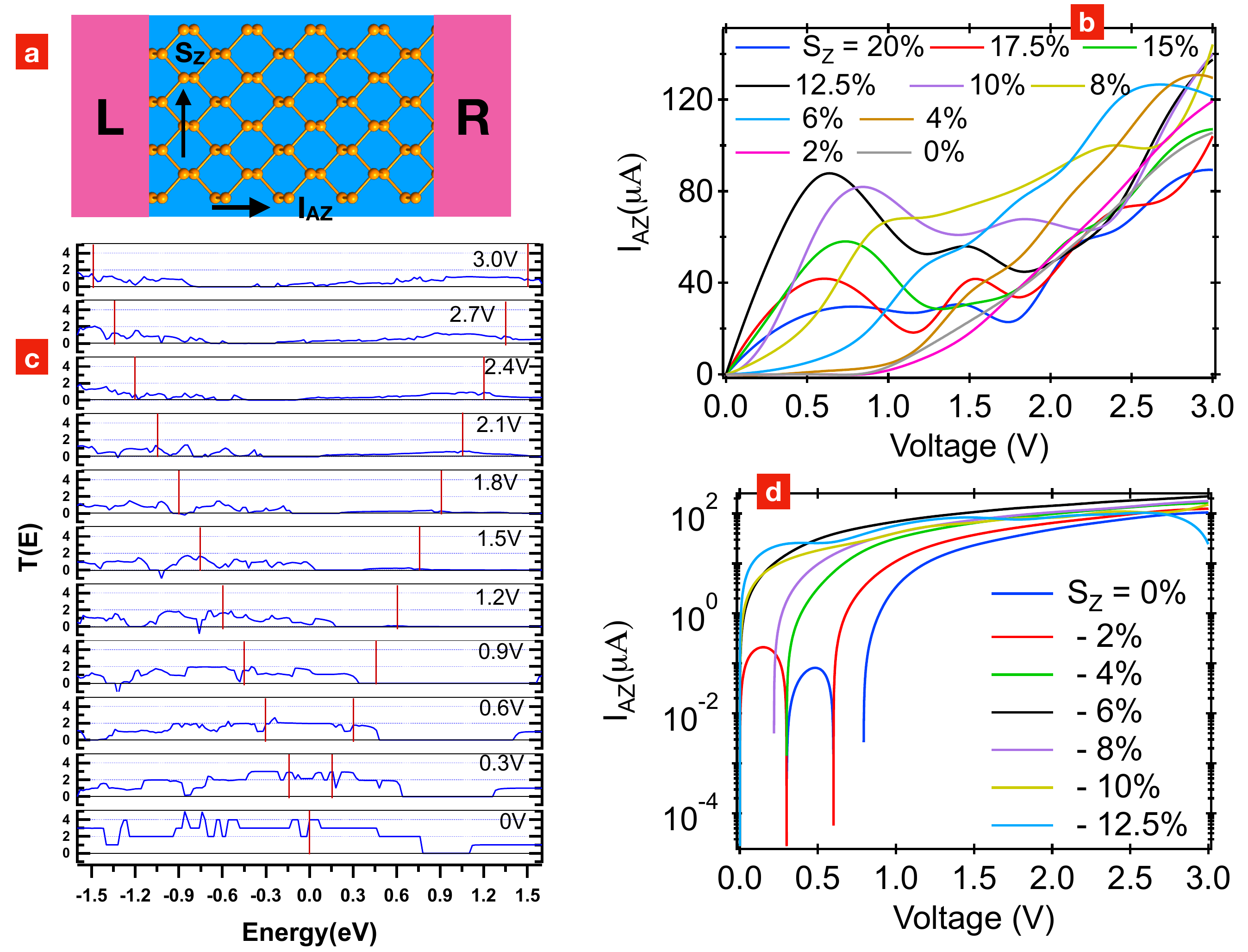}
\caption{{\small (a) Two-probe conduction geometry in Phosphorene for uniaxial strains along the zigzag direction and current measured along the armchair direction. (b) Corresponding Current-Voltage behaviour for various values of tensile strains and (d) compressive strain, (c) Transmission spectrum at various applied bias for a tensile strain of $S_Z$ = 12.5\%.}}
\label{fig.4}
\end{center}
\end{figure}  
The comparison made with $\Delta E$ can also be done from the Poisson's ratio or Young's modulus behaviour with similar strain conditions as well. Experimentally strain can be applied through a bending apparatus\cite{Conley2013}, nano-indentation\cite{Lee2008} or using flexible substrates. However, for applying preferential strain one has to identify the edge structure (zigzag or armchair) which can be done using a combination of optical and atomic force microscopy (AFM).

\begin{figure*}
\begin{center}
\includegraphics[width=18cm]{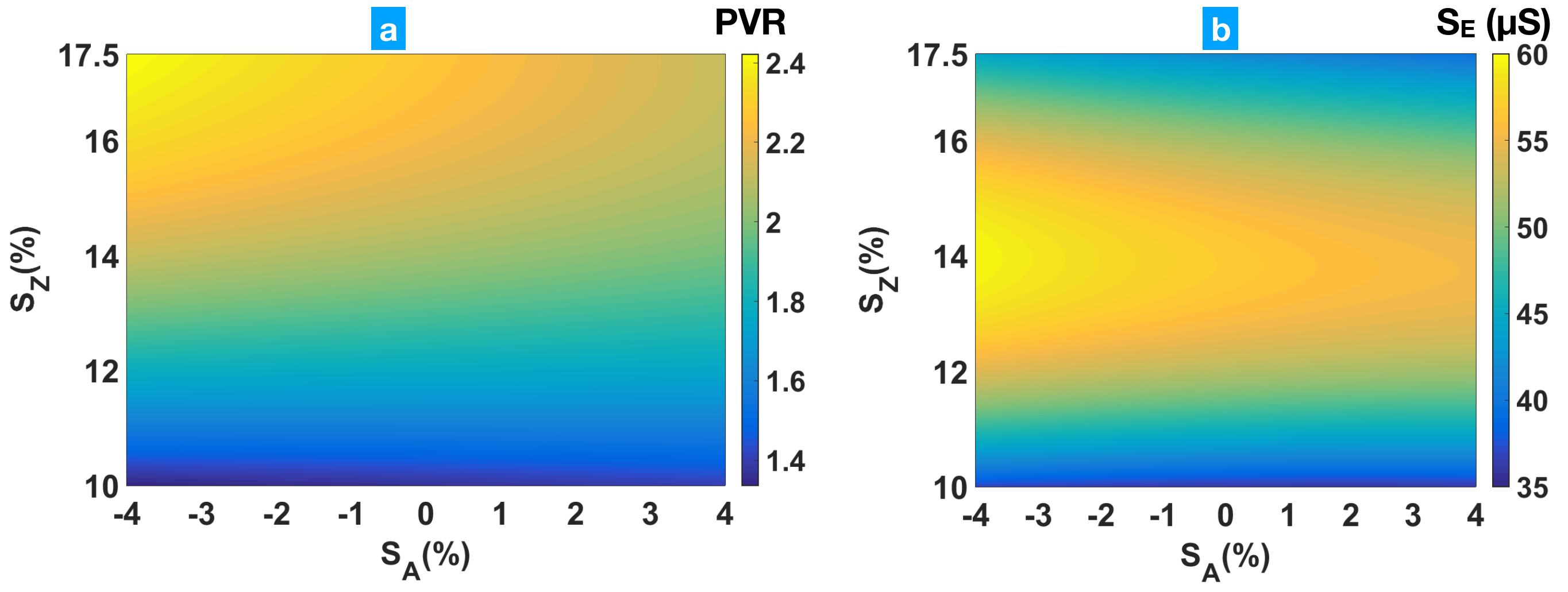}
\caption{{\small 2D plot of the (a) Peak to valley Ratio (PVR) and (b) Switching Efficiency ($S_E$}) as function of external strains applied along the armchair and zigzag directions. }
\label{fig.5}
\end{center}
\end{figure*} 
\subsection{Current-Voltage Characteristics}

For IV measurements, a finite sized central scattering region is connected between two semi-infinite electrodes marked by L (left) and R(right) [equivalent to Source (S) and Drain (D) in a Field-effect transistor (FET) geometry] as illustrated in Fig.~\ref{fig.3}(a-b). Considering the in-plane anisotropy of Phosphorene and its strain response, four different current components are  measured along the armchair and zigzag directions for two different strains, $S_A$ and $S_Z$. These are named as I$_{AZ}$, I$_{AA}$, I$_{ZA}$ and I$_{ZZ}$, where A and Z represents armchair and zigzag direction respectively. So,  I$_{AZ}$ (Fig.~\ref{fig.4}(a)) is the current measured along the Armchair (A) direction for external strain applied along the Zigzag (Z) direction. Similarly I$_{AA}$ is the current measured along the armchair direction for external strain ($S_A$) applied along the same direction as shown in Fig.~\ref{fig.3}(a).

\subsubsection{Parallel configuration for strain and current : }

I$_{AA}$ measured for different values of $S_A$ is illustrated in Fig.~\ref{fig.3}(c). At zero applied bias, the density of state of the two electrodes are identical leading to zero current. No significant increase in $I_{AA}$ is observed until $\sim$ 0.9V, which roughly corresponds to the size of the band gap of Phosphorene. Above V$>0.9V$, $I_{AA}$ increases almost exponentially with an increase in V, which is a signature of the gapped structure of the material in these strain regions. Between -4\% $ \le S_A\le $ +4\%, the shape of the IV curves do not show significant dispersions, which suggests that the electronic structure of the material mostly remain similar in this strain region.

In the $I_{ZZ} - V$ characteristics (Fig.~\ref{fig.3}(d)), zero current is observed between 0 - 0.2V bias. Above this, I$_{ZZ}$ increases sharply upto 1.4 V and a current plateau is observed between 1.4 - 1.5 V. Further increase in I$_{ZZ}$ continues upto $\sim$ 2.17V. At this voltage, a small peak in current is observed which is followed by a valley occurring around 2.45V. Above this, I$_{ZZ}$ increases sharply upto 3.0V. Such type of oscillations in the IV characteristics is reported here for the first time in Phosphorene. At high voltage, additional transport channels in phosphorene can contribute towards the sharp increase in I$_{ZZ}$. Similar to $I_{AA}$, no significant dispersions in $I_{ZZ}-V$ response is observed for various values of $S_Z$.

\subsubsection{Perpendicular configuration for strain and current : }

Very interesting features are observed in the $I_{AZ} - V$ patterns which is illustrated in Fig.~\ref{fig.4}. The IV curves in the TS regimes are shown in Fig.~\ref{fig.4}(b). In the low strain region ($S_Z = $0-4\%), the increase in $I_{AZ}$ is negligible at small applied bias and enhances rapidly thereafter, which is expected in a semiconductor with sizeable band gap. At higher values of $S_Z$, the gap size is reduced resulting in an increase in the $I_{AZ}$ at much smaller voltages. Systematic changes in the IV trends can be observed with an increase in strain, which starting at $S_Z = $10\% shows a distinct current peak at around 0.84 V after the initial increase in $I_{AZ}$ at low bias. At V$>$ 0.85V, $I_{AZ}$ started decreasing followed by a distinct valley (minimum in I$_{AZ}$) at 1.43 V, the subsequent secondary valley at around 2.21 V and sharp increase at higher voltages. This is a strong signature of current oscillations in Phosphorene which gets even more pronounced at $S_Z =$ 12.5\% and survives until $S_Z \sim$ 17.5\%. The oscillations are more prominent in the conductance behaviour as shown in Fig.~S1\cite{SI-Phosphorene-strain} for various values of TS. Unlike the commonly observed single peak and valley region in the IV-pattern, here the current/conductance goes through multiple oscillations which is unconventional for a 2D material.

The occurrence of such conductance oscillations can be explained using the transmission function T(E, V) calculated at different voltages as shown in Fig.~\ref{fig.4}(c) for $S_Z =$12.5\%. In between 0 - 0.63 V of bias range, the energy window of transmission increases almost linearly with increasing applied bias, resulting in a linear increase in current as well. Then a current peak is observed at 0.63V. Above 0.63 V, with an increase in the applied voltage current starts decreasing and this trend continues upto 1.84 V. By examining the T(E, V) patterns, it can be observed that with an increase in the voltage window, the energy window for transmission increases simultaneously which has been indicated by the red line(s) in Fig.~\ref{fig.4}(c) for different applied voltages. However, above 0.6V the peak value of T(E,V) starts decreasing and the energy range over which non-zero transmission occurs also gets reduced, resulting in only few T(E,V) peaks to survive at few energy values which continues upto 1.8 V. Thus integrating T(E,V) over the specified energy range gives an overall reduced transmission resulting in a lower value of I$_{AZ}$ in this voltage range. Above 1.84 V, with an increase in the energy window, the T(E,V) also starts increasing at various energies. Thus the integrated sum of transmission also enhances the current flowing between the leads which keeps continuing upto the maximum applied bias of 3 V. The conductance oscillations at other strains ($S_Z = $10\%, 15\% and 17.5\%) can be explained similarly using the changes of their respective T(E,V) at various applied voltages. 

However, under the application of CS, such oscillations were not observed as shown in Fig.~\ref{fig.4}(d). For low values of the CS, the material stays semiconducting and with an increase in CS, the voltage at which current starts flowing gets reduced. This follows a similar trend in the band gap evolution with strain as observed earlier from BS calculation. At higher values of the applied strain, the material turns metallic and it starts conducting even at very small applied bias near zero voltage.

\begin{figure}
\begin{center}
\includegraphics[width=9cm]{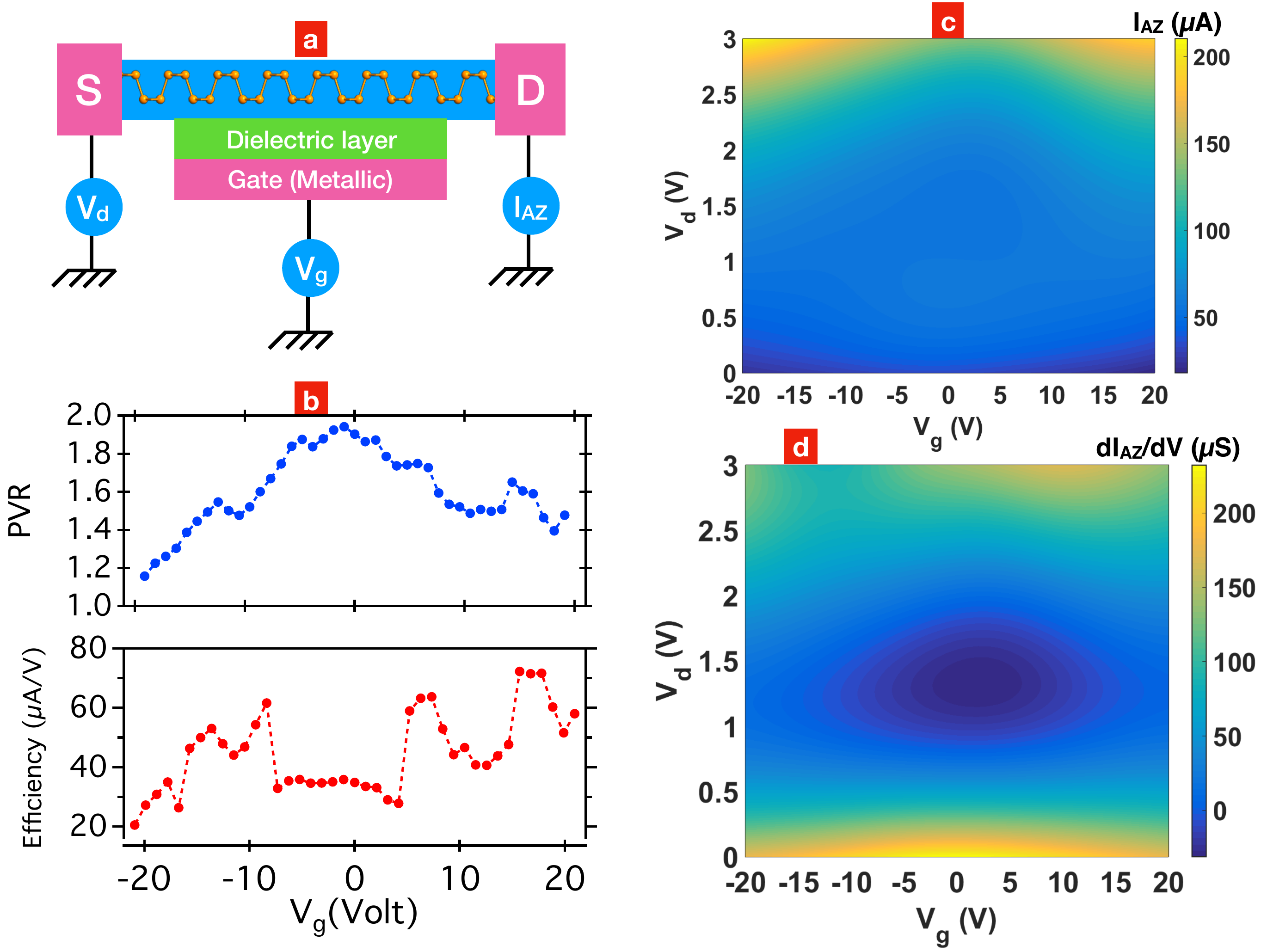}
\caption{{\small (a) Phosphorene layer based Field Effect Transistor with $S_Z$ = 12.5\%, $S_A$ = 0\%, (b) PVR and $S_E$ as function of the gate voltage, (c) Current-Voltage and (d) differential conductance response for various values of V$_g$ and V$_d$ projected on a 2D plane.}}
\label{fig.6}
\end{center}
\end{figure} 
 
The observation of such oscillatory differential conductance (and current) in unpassivated and strained Phosphorene is reported here for the first time. The strain tenability of these oscillations is also highly novel which can be used to design various switching devices by tuning the amplitudes of primary oscillations in desired regions. In order to quantify such current modulations relevant for device applications, two different parameters are used. The peak to valley ration (PVR) is defined as, 
\begin{equation}
\rm{PVR =  \Bigg\lvert \frac{I_{peak}}{I_{valley}} \Bigg\rvert}
\label{eqn.3} 
\end{equation}
and the switching efficiency is defined as,
\begin{equation}
\rm{S_E =\Bigg\lvert \frac{I_{peak}-I_{valley}}{V_{peak}-V_{valley}} \Bigg\rvert}
\label{eqn.4}
\end{equation}
where $\{I_{peak}, V_{peak}\}$ and $\{I_{valley}, V_{valley}\}$ are the current and voltages respectively in the peak and valley positions of the IV-curves. PVR is a measure of the limits of current(s) over which the ON/OFF operation can be performed safely between the peak and valley. The $S_E$ on the other hand is an estimation of how fast the material can be switched between the peak (ON) and valley (OFF) regions, thus relating to the speed of operation and fastness of switching.

In Fig.~\ref{fig.4}(b), significant modulations in the IV behaviour were observed for $S_Z =$ 10-17.5\% while $S_A = 0$. To get further insight about these oscillations, I$_{AZ}$-V characteristics were studied over this $S_Z$ range while the $S_A$ was varied between -4 to 4\%. The resulting PVR and $S_E$ estimated from these I$_{AZ}$-V curves are plotted in the form of 2D colour maps as illustrated in Fig.~\ref{fig.5}. The PVR in Fig.~\ref{fig.5}(a) has a maximum value of $\sim$ 2.4 occurring in the top left corner of the phase diagram. This shows that a combination of $S_Z > 0$ and $S_A < 0$ provides maximum PVR. For different values of $S_A$ and $S_Z$ considered here, PVR varies between 2.4 to 1.4, which indicates the robustness of these oscillations over significant strain variation in Phosphorene. Unlike the PVR behaviour, large value of $S_E$ is observed between $S_E =$ 12 - 16\% with a maximum of 60 $\mu$S at $S_Z$ = 14\% and $S_Z$ = -4\% as shown in Fig.~\ref{fig.5}(b). Combining Fig.~\ref{fig.5}(a) and Fig.~\ref{fig.5}(b), it is observed that to achieve a good PVR and $S_E$, preferred region of operation should be between $S_Z$ = 14-16\% with $S_A$ = -4\%. Thus strain tenability using a combination of both TS and CS is very much useful in finding the appropriate region of operation with preferred PVR and $S_E$ to achieve better switching and energy efficiency.

\paragraph*{\bf{The inclusion of Gate :\,\,}}

To understand the effect of a gate electrode on the observed conductance oscillations behaviour, the transport properties of strained Phosphorene ($S_Z$ = 12.5\%, $S_A$ = 0\%) was studied in a FET geometry as illustrated in Fig.~\ref{fig.6}(a). The metallic gate is connected through a dielectric layer to the Phosphorene channel region. The dielectric constant $\epsilon_r$ = 4.2 of the layer is similar to that of SiO$_2$, commonly used in FETs. The Source (S) and Drain (D) electrodes are also considered to be made of Phosphorene. This rules out the presence of barriers at the metal-semiconductor interfacial contacts and their effects on transport properties. For different values of $V_g$, I$_{AZ}$ and differential conductance (dI$_{AZ}$/dV) were estimated over a wide range of source-drain bias ($V_d$) as shown in the 2D colour plots Fig.~\ref{fig.6}(c-d). In Fig.~\ref{fig.6}(c), I$_{AZ}$ is symmetrically  distributed over the $V_d -V_g$ plane with its maximum at the two top corners around $V_d$ = 3V. For both high and low values of $V_g$, a large value of I$_{AZ}$ around $V_d$ = 3V suggests that a large transverse electric field can be used to create additional conduction pathways contributing to a large current flowing across the channel. The observation of large current ($\sim 200 \mu$A) at high values of the $V_g$ (= +20V, -20V) is an indication of the ambipolar nature of the material. This peak and valley feature are more pronounced in the differential conduction plot in Fig.~\ref{fig.6}(d). For most part of the $V_g$ range, the peak and valley areas are observed around the dark blue region in the 2D map. The largest slope of the I$_{AZ}-V$ curves was observed around $V_d = 0$V and 3V for, most values of $V_g$.

The PVR and $S_E$ estimated at different values of $V_g$ is shown in Fig.~\ref{fig.6}(b). The maximum in PVR was observed at $V_g$ = -1V with negative slopes on both sides towards $V_g$ = +20V and -20V. The difference between the maximum and minimum in PVR is 0.8, indicating that the characteristic oscillations can sustain over a significant V$_g$ swing. From an application point of view, a smaller $V_g$ is preferred to get a better PVR in this system. At high values of $V_g$, the phosphorene layer experiences a very high electric field from the gate as it is a short-channel device. It leads to an overall higher level of conduction in the valley regions and reduction of the PVR. $S_E$ on the other hand goes through a minimum between $V_g$ = -7V to +4V and starts increasing on both sides of this region with its maximum occurring at $V_g$ = 15V. Contrary to PVR, the $S_E$ shows significant improvement in the presence of $V_g$ with a maximum change of 37.9 $\mu$S ($> 100\%$) from the $V_g =0$V case. Combining them, a $V_g$ range between +15 V to +17V is found to get an optimum balance between these two parameters. We have also tested the FET in the lateral gate structure, where two gates were positioned on the two sides of the Phosphorene layer in its plane (Figure. S3\cite{SI-Phosphorene-strain}). However, no major changes in the PVR and $S_E$ were observed with the change of $V_g$. Thus in terms of tenability, a vertical gate configuration as indicated in Fig.~\ref{fig.6}(a) is preferred for nanoelectronic applications.

\paragraph*{\bf{Effect of Doping :\,\,}}

\begin{figure}
\begin{center}
\includegraphics[width=9cm]{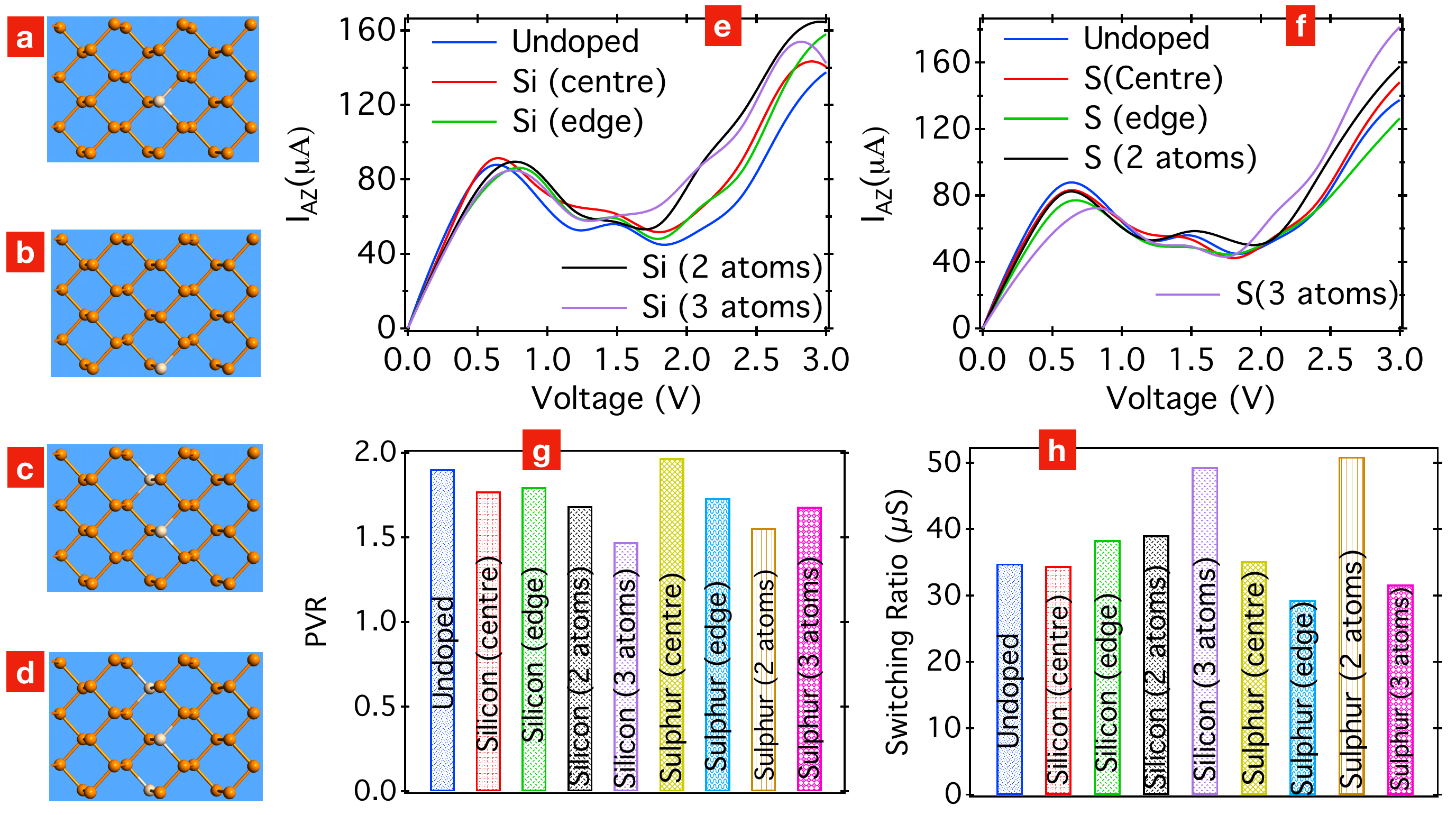}
\caption{{\small Various doping configurations in Phosphorene as (a) Centre (2.8\%), (b) Edge (2.8\%), (c) 2 atoms (5.6\%) and (d) 3 atoms (8.3\%). Current-Voltage relationship for these doping configurations with (e) Si and (b) S doping. (g) PVR and (h) $S_E$ for these doping configurations.}}
\label{fig.7}
\end{center}
\end{figure}

To check the effect of doping on the conductance oscillations of strained Phosphorene ($S_Z$ = 12.5\%, $S_A$ = 0\%), IV measurements were carried out under different doping conditions. This includes doping using both $p$ (Silicon) and $n$ (Sulphur) type dopants at specific sites and at varied concentration as shown in Fig.~\ref{fig.7}(a-d). The corresponding IV characteristics is illustrated in Fig.~\ref{fig.7}(e-f). Both for Si and S doping, the overall shape of the IV curves do not change significantly with a change of the doping configurations and the  current oscillations are observed in all the cases. There is no significant dispersions observed in $I_{AZ}$ for the entire voltage range which suggest that the oscillations are pretty robust against the nature, location and concentration of doping in this case.  

PVR estimated for these configurations is shown in Fig.~\ref{fig.7}(g). The maximum of PVR is observed when Sulphur is doped at the centre and the lowest for doped Silicon (3 atoms). Unlike this, the $S_E$ shows significant improvement under both $p/n$-type of doping compared to its undoped counterpart. Maximum of $S_E$ is observed for S doping (2 atoms) with $S_E =$ 50.9 $\mu$S which is almost comparable to Si-doping (3 atoms) with 49.4 $\mu$S. Both these values are more than 40\% higher than the $S_E$ obtained from the undoped configuration. This trend clearly shows that a higher level of doping can help achieving better functionality in this strained structure. Combing the PVR and $S_E$, the preferred configuration for optimum performances are : (i) Si (edge), (ii) Si (2 atoms) and (iii) S (2 atoms) which corresponds to a doping percentage between 2.8\% to 5.6\%. 

\begin{figure}
\begin{center}
\includegraphics[width=9cm]{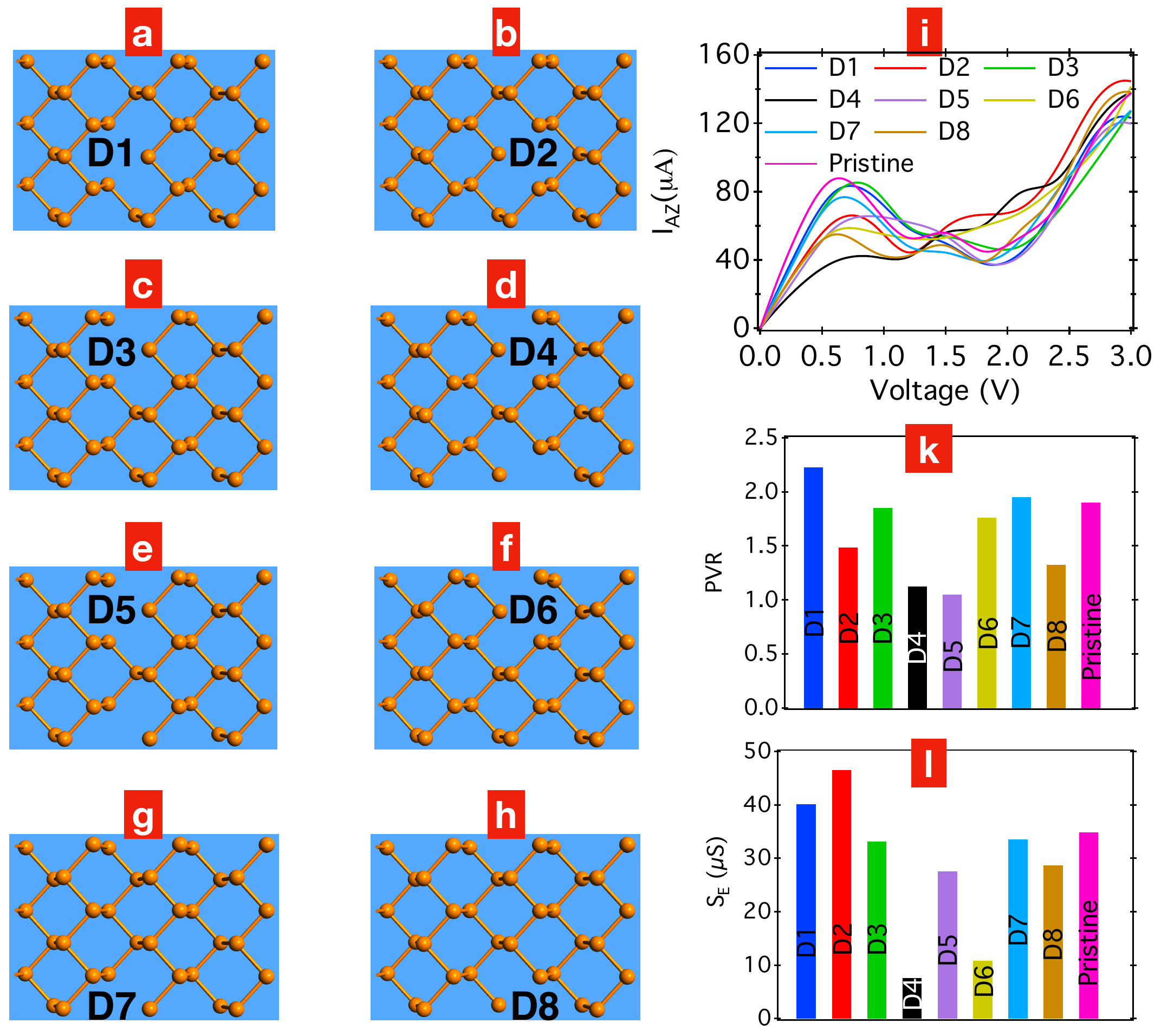}
\caption{{\small Monolayer Phosphorene in various defect configurations named as (a) D1, (b) D2, (c) D3, (d) D4, (e) D5, (f) D6, (g) D7, (h) D8. (i) Current-Voltage relationship of Phosphorene in these defect configurations. (j) PVR and (k) $S_E$ for these defect configurations.}}
\label{fig.8}
\end{center}
\end{figure}

\paragraph*{\bf{Influence of Defect :\,\,}}
We have also checked the influence of various types of defects in strained Phosphorene ($S_Z$ = 12.5\%, $S_A$ = 0\%) on the IV-curves as illustrated in Fig.~\ref{fig.8}. In Fig.~\ref{fig.8}(a-h), eight distinct defect configurations are shown namely, D1 - D8 based on the locations of defects at the centre or edge of the layer. The corresponding IV-curves for these configurations are displayed in Fig.~\ref{fig.8}(i), where noticeable dispersions between different defect configurations can be clearly seen. For example, between D4 and D7 a difference in current $\sim40\mu$A is significant for a small sized system. In the strained configuration, the metallic nature of the material provides transmission pathways with ample charge carriers available for conduction resulting in robust  IV-characteristic obtained in such defect configurations.

The rate of enhancement of $I_{AZ}$ from V=0 to the peak region(s) is different for D1 - D8 and this affects the peak current observed. The current oscillations are prominent in all these cases and double valley structure is present for some of them. The PVR and $S_E$ estimated from these curves are plotted in Fig.~\ref{fig.8}(k) and Fig.~\ref{fig.8}(l) respectively. The highest value of PVR was observed for D1, although D3, D6 and D7 configurations show a PVR comparable to that of the pristine counterpart. D1 is also found to be the most stable defect configuration in terms of energy consideration.

Similarly significant dispersions are noticeable in $S_E$ pattern, which has a maximum for D2 while D3, D7 configurations show a comparable $S_E$ to that of the defect-less material. The lowest values of PVR and $S_E$ are observed for D4 and D5 configurations as in both these cases, the numbers of atoms removed only allows a tiny part of the sample to conduct currents between the two electrodes. This reduces the current at the peak and valley regions of the respective curves. Comparing between various defect configurations, D1 seems to be the most preferred configuration in terms of improved PVR ($\sim17\%$) and $S_E$ ($\sim15\%$) higher than the pristine counterpart. D7, D3 and D2 configurations also provide a good balance between PVR and $S_E$ with comparable numbers to that of the pristine material while D4 and D5 are the least preferred configurations. As defects are common in most 2D materials, hence the results and methodology explained here can be useful in tuning the conductance oscillations if prior information about the defect types is known about the material. 

We also studied the effect of edge-passivation on the oscillatory behaviour of the conductance. However, no significant changes in the PVR and $S_E$ were observed for edge passivation with C and H atoms on the strained Phosphorene. Doping can introduce similar features in IV characteristic as observed for graphene \cite{Dragoman2007, Nguyen2011,Wu2012} and MoS$_2$ \cite{Sengupta2013}, but such changes are irreversible. Unlike such permanent changes, here the robust conductance oscillations can be tuned reversibly by strain alone, without any doping. This is advantageous for employing the phenomena for control of charge or spin information to create advanced fast switches, high speed inverted transistors, high-frequency THz oscillators, frequency synthesizers etc. Furthermore, engineering contacts and controlling contact transport mechanism\cite{Venkata2D} could lead to additional flexibility for controlled applications. Because of the growing prospects in Phosphorene, these new  possibilities can make Phosphorene a core material to advance present day 2D electronics radically.

\section{Conclusion}

In this work we have performed First-principles based DFT investigation of the electronic transport properties of monolayer Phosphorene under the influence of various uniaxial and biaxial strains, applied along its two principle armchair and zigzag direction. We observe unconventional conductance oscillation in the current-voltage characteristics by the application of simple strain. The high amplitude oscillatory behaviour can be characterised by a switching efficiency as high as $\sim80\mu S$. We observe that such behaviour can be modulated faithfully by applying a gate voltage and the oscillatory feature is robust regardless of the defects and doping. To obtain better functionality, we identified the optimum configuration of external stimulus that can lead to high PVR and Switching efficiency. Considering the fact that modern nanotechnological tools allow us to create sub-10 nm structures, the feasibility of practical studies in this direction is also high. Our results add new insights into the strain-dependent physics of Phosphorene, propose new avenues for stain dependent devices in nanoelectronic circuits and present new prospects for future experimental investigations.

\section*{Acknowledgments}

This work was financially supported by Department of Science and Technology through the  INSPIRE scheme (Ref: DST/INSPIRE/04/2015/003087). SJR sincerely acknowledges the support provided by Indian Institute of Technology Patna. G. Nirala and S. Kumari are acknowledged for assistance with the initial analysis. MVK acknowledges Swedish Research Council VR Starting grant No.2016-03278.


\footnotesize{
\bibliography{Bibliography-ATK} 
\bibliographystyle{rsc} 
}

\end{document}